\documentclass[journal,a4paper]{IEEEtran}
\usepackage[T1]{fontenc}% optional T1 font encoding
\usepackage{layouts}
\usepackage{cite}
\usepackage[pdftex,colorlinks=true,bookmarks=true,citecolor=blue,urlcolor=blue]{hyperref} %pdflatex

\ifCLASSINFOpdf
  \usepackage[pdftex]{graphicx}
   \graphicspath{{./figs/}}
  \DeclareGraphicsExtensions{.pdf,.jpeg,.png}
\else
\fi
\usepackage{amsmath}
\usepackage{amsthm}
\usepackage[cmintegrals]{newtxmath}
\usepackage{bm}
\usepackage{algorithmic}
\usepackage{array}
\ifCLASSOPTIONcompsoc
  \usepackage[caption=false,font=normalsize,labelfont=sf,textfont=sf]{subfig}
\else
  \usepackage[caption=false,font=footnotesize]{subfig}
\fi
\usepackage{mathrsfs}  
\usepackage{amsbsy}

\newcommand{\field}[1]{\mathbb{#1}}
\newcommand{\sdots}{\reflectbox{$\ddots$}}

\newcommand{\tp}{\intercal}% transpose operation

\newcommand{\bigO}[1]{\mathop{\mathcal{O}}\left(#1\right)}

\newcommand{\vv}[1]{\mathbf{#1}}
\newcommand{\vs}[1]{\boldsymbol{#1}}

\newcommand{\OP}[1]{\mathscr{#1}}

\DeclareMathOperator{\sech}{sech}

\usepackage{enumitem}

\begin{document}
\title{A Fast Chebyshev Spectral Method for Nonlinear Fourier Transform}

%\author{Michael~Shell,~\IEEEmembership{Member,~IEEE,}
%        John~Doe,~\IEEEmembership{Fellow,~OSA,}
%        and~Jane~Doe,~\IEEEmembership{Life~Fellow,~IEEE}% <-this % stops a space

\author{Vishal Vaibhav% <-this % stops a space
\thanks{Email:~\tt{vishal.vaibhav@gmail.com}}% <-this % stops a space
}
\maketitle

\begin{abstract}
In this letter, we present a fast and well-conditioned spectral method based on 
the Chebyshev polynomials for computing the continuous part of the nonlinear 
Fourier spectrum. The algorithm achieves a complexity of 
$\bigO{N_{\text{iter.}}N\log N}$ per spectral node for $N$ samples of 
the signal at the Chebyshev nodes where $N_{\text{iter.}}$ is the number of 
iterations of the biconjugate gradient stabilized method.
\end{abstract}

%\begin{IEEEkeywords}
%\end{IEEEkeywords}

% \fi
\IEEEpeerreviewmaketitle

\section{Introduction}
This paper considers the Zakharov and Shabat (ZS)~\cite{ZS1972} scattering 
problem which forms the basis for defining a nonlinear generalization of 
the conventional Fourier transform dubbed as the 
\emph{nonlinear Fourier transform} (NFT). In an 
optical fiber communication system the nonlinear Fourier (NF) spectrum 
offers a novel way of encoding information in optical pulses where the nonlinear effects 
are adequately taken into account as opposed to being treated as a source of 
signal distortion~\cite{Yousefi2014compact,TPLWFK2017}. One of the challenges 
that has emerged in realizing these ideas is the development of a fast and 
well-conditioned NFT algorithm that can offer spectral accuracy at low 
complexity. Such an algorithm would prove extremely useful for system design 
and benchmarking. Currently, there are primarily two successful 
approaches proposed in the literature for computing the continuous 
NF spectrum which are capable of achieving algebraic orders convergence 
at quasilinear complexity: (a) the integrating factor (IF) based exponential 
integrators~\cite{V2017INFT1,V2018BL,V2018LPT,V2019LPT} (b) exponential time 
differencing (ETD) method based exponential 
integrators~\cite{V2019CNFT}. Note that while the IF schemes uses fast 
polynomial arithmetic in the monomial basis, the ETD schemes 
use fast polynomial arithmetic in the Chebyshev basis. For the inverse 
transform, a sampling series based approach for computing the 
``radiative'' part has been proposed in~\cite{V2019BL1} which achieves spectral 
accuracy at quasilinear complexity per sample of the signal. In this paper, we propose 
a spectral method for direct NFT that achieves quasilinear complexity per 
sample of the continuous spectrum. The signal in this case must be sampled at 
the so-called Chebyshev--Gauss--Lobatto nodes.

Introducing the ``local'' scattering coefficients $a(t;\zeta)$ and 
$b(t;\zeta)$ such that 
$\vs{\phi}(t;\zeta)=(a(t;\zeta)e^{-i\zeta t}, b(t;\zeta)e^{i\zeta t})^{\tp}$, 
the ZS scattering problem can be written as 
$\partial_{t}a(t;\zeta)=q(t)b(t;\zeta)e^{2i\zeta t}$ and 
$\partial_{t}b(t;\zeta)=r(t)a(t;\zeta) e^{-2i\zeta t}$. Let 
the scattering potential $q(t)$ (with $r(t)=\alpha q^*(t)$ where 
$\alpha\in\{+1,-1\}$) be supported in $\field{I}=[-1,1]$. Specializing to the real 
line $\zeta=\xi\in\field{R}$, the 
initial conditions for the Jost solution $\vs{\phi}$ 
are: $a(-1;\xi) = 1$ and $b(-1;\xi) = 0$. The scattering coefficients 
$\mathfrak{a}$ and $\mathfrak{b}$ are given by $\mathfrak{a}(\xi)=a(+1;\xi)$ 
and $\mathfrak{b}(\xi)=b(+1;\xi)$ so that the reflection coefficients is 
$\rho=\mathfrak{b}/\mathfrak{a}$. Let $g(t;\xi)=q(t)e^{+2i\xi t}$ and 
$h(t;\xi)=\alpha g^*(t;\xi)$. In the following, we describe a 
numerical scheme based on the Chebyshev polynomials to solve the coupled 
Volterra integral equations given by 
\begin{equation}\label{eq:Volterra-ZS}
a(t)=1+\int_{-1}^tg(s)b(s)ds,\quad
b(t)=\int_{-1}^th(s)a(s)ds,
\end{equation} 
derived from the ZS problem using the aforementioned initial conditions where 
the dependence on $\xi$ is suppressed for 
the sake of brevity of presentation. We refer the reader to~\cite{V2018TL} for the 
nonlinear Fourier analysis of time-limited signals.

\section{The numerical scheme}
In this section, we describe the numerical scheme and carry out the theoretical 
analysis of its convergence and stability. Let us set the following notations:
$\field{N}=\{1,2,\ldots\}$ and $\field{N}_0=\{0\}\cup\field{N}$ (the set of whole numbers), 
$\ell^{p}$ ($1\leq p\leq\infty$) denotes the discrete Lebesgue spaces with norm 
$\|\vv{C}\|_p=|\sum_{n=0}^{\infty}|C_n|^p|^{1/p},\,1\leq p<\infty,$ and
$\|\vv{C}\|_{\infty}=\sup\{C_0,C_1,\ldots\}$.

The first step is to obtain a discrete 
representation of an integral operator in the Chebyshev basis using the 
relations
\begin{equation*}
\begin{split}
&\int_{-1}^tT_0(s)ds=T_1(t)+T_0(t),\quad
\int_{-1}^tT_1(s)ds=\frac{T_{2}(t)}{4}-\frac{T_{0}(t)}{4},\\
&\int_{-1}^tT_n(s)ds=\frac{T_{n+1}(t)}{2(n+1)}-\frac{T_{n-1}(t)}{2(n-1)}
-\frac{(-1)^{n}T_0(t)}{n^2-1},\quad n>1.
\end{split}
\end{equation*}
Let $c(t)=\sum_{n=0}^{\infty}C_nT_n(t)$ and $d(t)=\OP{K}[c](t)$ where 
$\OP{K}[c](t)=\int_{-1}^tc(s)ds$, we have
\begin{multline}
d(t)=\left[C_0-\frac{1}{4}C_1-\sum_{n=2}^{\infty}\frac{(-1)^{n}C_n}{n^2-1}\right]T_0(t)\\
+\left[C_0-\frac{1}{2}C_2\right]T_1(t)+
\sum_{n=2}^{\infty}\frac{1}{2n}\left[C_{n-1}-C_{n+1}\right]T_n(t).
\end{multline}
Now setting $d(t)=\sum_{n=0}^{\infty}D_nT_n(t)$, and, introducing the infinite 
dimensional vectors $\vv{C}=(C_0,C_1,\ldots)^{\tp}$ and 
$\vv{D}=(D_0,D_1,\ldots)^{\tp}$, we have
\begin{equation}
\vv{D}=
\begin{pmatrix}
 1 &-\frac{1}{4} &  -\frac{1}{3} & +\frac{1}{8} & -\frac{1}{15}  & \ldots\\
 1 &          0  & -\frac{1}{2}&  &   & \\
   &   \frac{1}{4}&          0 & -\frac{1}{4} &            & \\
   &             &  \frac{1}{6} &      0   &-\frac{1}{6}   &\\
   &             &             & \ddots       &\ddots   &\ddots
\end{pmatrix}
\vv{C}\equiv \mathcal{K}\vv{C}.
\end{equation}
The next step in the discretization of~\eqref{eq:Volterra-ZS} involves expanding the scattering 
potentials in the Chebyshev basis. Let 
$g(t) = \sum_{n=0}^{\infty}G_nT_n(t)$ and $h(t) = \sum_{n=0}^{\infty}H_nT_n(t)$ 
where $H_n=\alpha G^*_n$. A truncated expansion upto $N$ terms can be accomplished by 
sampling the potentials at the Chebyshev--Gauss--Lobatto nodes given by 
$t_n = -\cos[n\pi/(N-1)],\,n=0,1,\ldots N-1$ and carrying out discrete Chebyshev transform 
which can be implemented using an FFT of size $2(N-1)$~\cite{G2011}. Now, our final goal is to 
obtain an expansion of the local scattering coefficients in the Chebyshev basis: To this end, let
$a(t)= \sum_{n=0}^{\infty}A_nT_n(t)$ and $b(t)= \sum_{n=0}^{\infty}B_nT_n(t)$ where $A_n$ and 
$B_n$ are to be determined (for fixed value of $\xi$). The last ingredient 
needed in the discretization of~\eqref{eq:Volterra-ZS} are the products $h(t)a(t)$ and 
$g(t)b(t)$ which must be represented as a linear operation on the unknown coefficient vectors
$\vv{A}=(A_0,A_1,\ldots)^{\tp}$ and $\vv{B}=(B_0,B_1,\ldots)^{\tp}$. To this end, let
$h(t)a(t)= \sum_{l=0}^{\infty}C_lT_l(t)$ and $g(t)b(t)= \sum_{l=0}^{\infty}D_lT_l(t)$;
then, it follows that $2C_0 = 2H_0A_0 +\sum_{k=1}^{\infty}H_kA_k$,
$2D_0 = 2G_0B_0 +\sum_{k=1}^{\infty}G_kB_k$ and
\begin{equation}
\begin{split}
2C_l &=\sum_{k=0}^{l-1}H_{l-k}A_k
+2H_{0}A_{l}+\sum_{k=1}^{\infty}H_{k}A_{k+l}
+\sum_{k=0}^{\infty}H_{k+l}A_k,\\
2D_l &=\sum_{k=0}^{l-1}G_{l-k}B_k
+2G_0B_l+\sum_{k=1}^{\infty}G_{k}B_{l+k}
+\sum_{k=0}^{\infty}G_{k+l}B_k.
\end{split}
\end{equation}
for $l\in\field{N}$. Following Olver and Townsend~\cite{OT2013}, these relations 
define the operator $\mathcal{M}[\vv{G}]$, which comprises a T\"oplitz and an 
almost Hankel matrix given by
\begin{equation}
2\mathcal{M}[\vv{G}]=
\begin{pmatrix}
 2G_0  & G_1 &  G_2 &\cdots\\
  G_1 & 2G_0 &  G_1 &\ddots\\
  G_2 &  G_1 & 2G_0 &\ddots\\
\vdots&\ddots&\ddots&\ddots
\end{pmatrix}\\
+
\begin{pmatrix}
   0  &  0   &   0  &\cdots\\
  G_1 &  G_2 & G_3  &\sdots\\
  G_2 &  G_3 & G_4  &\sdots\\
\vdots&\sdots&\sdots&\sdots
\end{pmatrix},
\end{equation}
Similarly, the representation of the operator $\mathcal{M}[\vv{H}]$ follows 
from the same convention where we recall $\vv{H}=\alpha\vv{G}^*$. Setting 
$\Lambda=\mathcal{K}\mathcal{M}[\vv{G}]$, the discrete version 
of~\eqref{eq:Volterra-ZS} can be stated as
\begin{equation}\label{eq:Volterra-ZS-discrete}
\begin{pmatrix}
I & -\Lambda\\
-\alpha\Lambda^* & I
\end{pmatrix}
\begin{pmatrix}
\vv{A}\\
\vv{B}
\end{pmatrix}=
\mathcal{S}
\begin{pmatrix}
\vv{A}\\
\vv{B}
\end{pmatrix}=
\begin{pmatrix}
\vv{E}_0\\
\vv{0}
\end{pmatrix},
\end{equation}
where $I$ is the identity matrix, $\vv{E}_0=(1,0,\ldots)^{\tp}$ and 
$\vv{0}=(0,0,\ldots)^{\tp}$. Noting that $\vv{B}=\alpha\Lambda^*\vv{A}$ and 
setting $\Gamma = \alpha\Lambda\Lambda^*$, we 
have $\left(I-\Gamma\right)\vv{A}=\vv{E}_0$. Before we attempt to solve this equation, let 
us determine the conditions under which a solution exists. To this end, let us 
show that the infinite matrices $\mathcal{K}$ and $\mathcal{M}[\vv{G}]$ 
(or, equivalently, $\mathcal{M}[\vv{H}]$) form bounded 
linear operators on $\ell^{\infty}$: Let $\vv{C}\in\ell^{\infty}$ and consider 
$\vv{D}=\mathcal{K}\vv{C}$, then
\begin{multline}
\|\vv{D}\|_{\infty}=\sup\biggl\{
\left|C_0-\frac{1}{4}C_1-\sum_{n=2}^{\infty}\frac{(-1)^{n}C_n}{n^2-1}\right|,
\left|C_0-\frac{1}{2}C_2\right|,\\
\frac{1}{4}\left|C_{1}-C_{3}\right|,
\frac{1}{6}\left|C_{2}-C_{4}\right|,
\ldots\frac{1}{2n}\left|C_{n-1}-C_{n+1}\right|,\ldots\biggl\}.
\end{multline}
The dominant term, which is clearly $D_0$, satisfies the estimate 
$|D_0|\leq ({7}/{4})\|\vv{C}\|_{\infty}$
%\begin{equation*}
%\left|C_0-\frac{1}{4}C_1-\sum_{n=2}^{\infty}\frac{(-1)^{n}C_n}{n^2-1}\right|
%\leq \frac{7}{4}\|\vv{C}\|_{\infty},
%\end{equation*}
so that $\|\vv{D}\|_{\infty}\leq (7/4)\|\vv{C}\|_{\infty}$ which yields 
$\|\mathcal{K}\|_{\infty}\leq 7/4$.
Next, assuming that $\vv{G}\in\ell^1$ (which corresponds to the fact that 
$\partial_tg$ is absolutely continuous on $\field{I}$~\cite{2008T}), we have
$2|D_0|\leq\left(|G_0| +\|\vv{G}\|_{1}\right)\|\vv{C}\|_{\infty}$ and 
$|D_l|\leq\|\vv{G}\|_{1}\|\vv{C}\|_{\infty}$ for $l\in\field{N}$
so that $\|\vv{D}\|_{\infty}\leq\|\vv{G}\|_{1}\|\vv{C}\|_{\infty}$ which 
yields $\|\mathcal{M}[\vv{G}]\|_{\infty}\leq\|\vv{G}\|_1$.

Next, we would like to show that the inverse of $(I-\Gamma)$ is bounded on 
$\ell^{\infty}$. To this end, let $\vv{E}_n$ denote the $n$-th 
column of $I$; then, for any arbitrary 
$\vv{C}\in\ell^{\infty}$, we have $\vv{C}=\sum_{n=0}^{\infty}C_n\vv{E}_n$. Therefore, 
it suffices to show that $\|\mathcal{R}\vv{E}_n\|_{\infty}$ is bounded 
for all $n\in\field{N}_0$ where 
$\mathcal{R}$ the resolvent of $(I-\Gamma)$ defined by
$\mathcal{R}=\sum_{l=1}^{\infty}\Gamma^l$ so that 
$(I-\Gamma)^{-1}=I+\mathcal{R}$. Consider $\Gamma\vv{E}_n$; by definition
\begin{multline}
(\Gamma\vv{E}_n)_m= \frac{2}{\pi(1+\delta_{0m})}\int_{-1}^1dt\frac{T_m(t)}{\sqrt{1-t^2}}\\
\times\left(\int^t_{-1}ds_2 g(s_2;\xi)\int^{s_2}_{-1}ds_1 h(s_1;\xi)T_n(s_1)\right).
\end{multline}
Let $\chi(t)=\int_{-1}^t|q(s)|ds$. Following~\cite{V2018TL} and taking into 
account the definition of the Chebyshev coefficients together with the 
property $|T_n(t)|\leq1,\,n\in\field{N}_0$, we have
\begin{align*}
&\|\Gamma\vv{E}_n\|_{\infty}
\leq\frac{2}{\pi}\int_{-1}^1\frac{dt}{\sqrt{1-t^2}}\\
&\quad\times\left(\int_{-1}^tds_2|q(s_2)|
\int_{-1}^{s_2}ds_1|q(s_1)|\right)
\leq2\frac{[\chi(1)]^2}{1\cdot2},\\
%\end{multline*}
%Similarly,
%\begin{multline*}
&\|\Gamma^2\vv{E}_n\|_{\infty}
\leq\frac{2}{\pi}\int_{-1}^1\frac{dt}{\sqrt{1-t^2}}\\
&\quad\times\left(\int_{-1}^tds_4|q(s_4)|
\int_{-1}^{s_4}ds_3|q(s_3)|\frac{[\chi(s_3)]^2}{2!}\right)
\leq2\frac{[\chi(1)]^4}{4!}.
\end{align*}
It is possible to conclude that $\|\Gamma^l\vv{E}_n\|_{\infty}\leq2{[\chi(1)]^{2l}}/{(2l)!}$ for 
$l\in\field{N}$ continuing as above yielding the estimate
\begin{equation}
\|(I+\mathcal{R})\vv{E}_n\|_{\infty}\leq1+2\sum_{l=1}^{\infty}\frac{[\chi(1)]^{2l}}{(2l)!}
=2\cosh[\chi(1)]-1.
\end{equation}
From here, it also follows that $\|I+\mathcal{R}\|_{\infty}\leq(2\cosh[\chi(1)]-1)$. 

\begin{figure}[h!]
\centering
\includegraphics[scale=1]{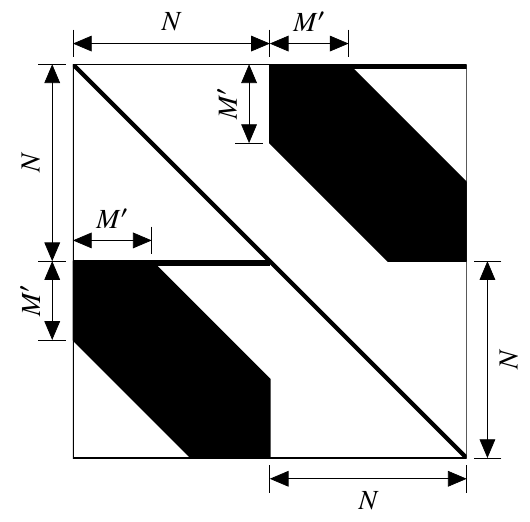}
\caption{\label{fig:spy-mat}The figure depicts the sparsity structure of 
the truncated version $\mathcal{S}_N\in\field{C}^{2N\times 2N}$ of the 
$2\times2$ block matrix $\mathcal{S}$ defined 
in~\eqref{eq:Volterra-ZS-discrete}. Here $M'=M+1$ where $M$ is 
the number of Chebyshev polynomials used for the potential function 
$g(t)$. The matrix $\mathcal{S}_N$ is banded with exactly two filled 
rows each corresponding to the filled top row of $\Lambda_N$.}
\end{figure}
\begin{figure*}[th!]
\centering
\includegraphics[scale=1]{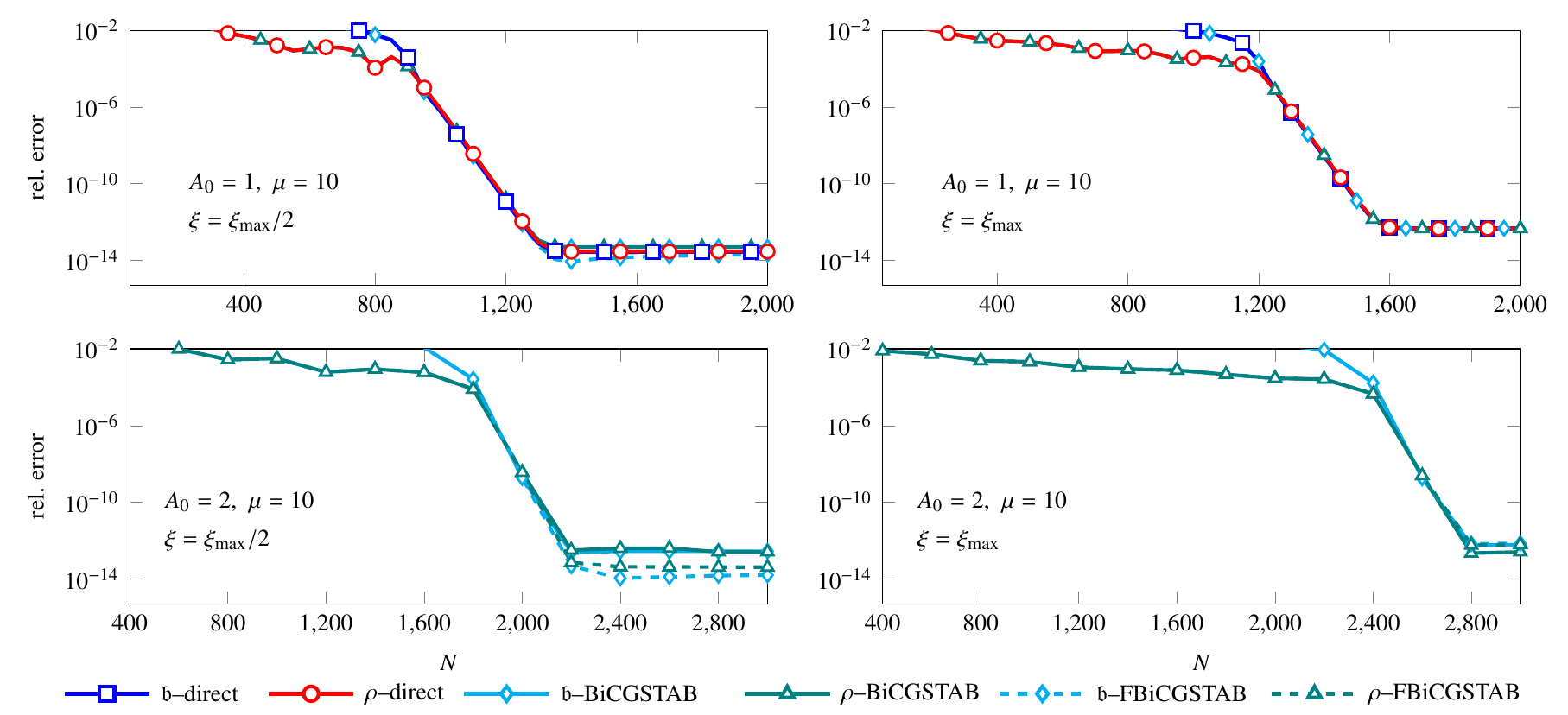}
\caption{\label{fig:convg-all}The figure shows the convergence analysis for the proposed 
algorithms using a chirped hyperbolic potential.}
\end{figure*}

The numerical algorithm is now obtained by truncating the Chebyshev expansion 
for $g(t)$ and, equivalently, $h(t)$ to $M$ terms while truncating $\Lambda$ to 
an $N\times N$ matrix where $N\geq2M$ so that the discrete system in reads as 
$\left(I_N-\Gamma_N\right)\vv{A}_N=(\vv{E}_0)_N$ where $\vv{A}_N\in\field{C}^N$ and 
$\Gamma_N = \alpha\Lambda_N\Lambda_N^*\in\field{C}^{N\times N}$.

The sparsity structure of the original linear system in~\eqref{eq:Volterra-ZS-discrete} is 
depicted in Fig.~\ref{fig:spy-mat}. Let $\mathcal{R}_N$ denote the resolvent of $(I_N-\Gamma_N)$; then, 
$\|I_N+\mathcal{R}_N\|_{\infty}\leq (2\cosh[\chi_M(1)]-1)$ where 
$\chi_M(t)=\int_{-1}^t|q_M(s)|ds$ ($q_M(s)$ being the $M$--term Chebyshev 
approximation to $q(s)$). 

In order to study the convergence behavior of the 
numerical algorithm, we consider 
the error $\|\vv{A}-\vv{A}_N\|_{\infty}$ where, with slight abuse of notation, we 
have assumed that $\vv{A}_N$ is infinite dimensional by setting the additional entries 
to $0$. We also extend the matrix $\Gamma_N$ to an infinite dimensional matrix in 
the same fashion. The error can be estimated from the relation 
$(I-\Gamma)(\vv{A}-\vv{A}_N) = (\Gamma-\Gamma_N)\vv{A}_N$ which yields
\begin{equation}
\begin{split}
\|\vv{A}-\vv{A}_N\|_{\infty}
&\leq\|I+\mathcal{R}\|_{\infty}\|\Gamma-\Gamma_N\|_{\infty}\|\vv{A}_N\|_{\infty}\\
&\leq\|I+\mathcal{R}\|_{\infty}\|I+\mathcal{R}_N\|_{\infty}\|\Gamma-\Gamma_N\|_{\infty}.
\end{split}
\end{equation}
The convergence of the numerical scheme, therefore, depends on $\|\Gamma-\Gamma_N\|_{\infty}$ 
which can be estimated as follows:
\begin{equation}
\|\Gamma-\Gamma_N\|_{\infty}
\leq2\|\mathcal{K}\|^2_{\infty}\|\mathcal{M}[\vv{G}]\|_{\infty}\sum_{l=M+1}^{\infty}|G_l|.
\end{equation}
If the first $k-1$ derivatives of $g(t)$ are absolutely continuous, then 
$\sum_{l=M+1}^{\infty}|G_l|=\bigO{M^{-k}}$~\cite{2008T}. Using the estimates obtained 
so far, an estimate for the condition number $\kappa$ of the matrix $(I_N-\Gamma_N)$ in the 
$\ell^{\infty}$--norm works out to be
$\kappa\leq2\cosh(\|\vv{Q}\|_1)\left(1+\|\mathcal{K}\|^2_{\infty}\|\vv{G}\|^2_1\right)$
which guarantees that the numerical scheme remains well-conditioned for all $N$.
%\begin{equation}
%\begin{split}
%\kappa(I_N-\Gamma_N)
%&=\|(I_N-\Gamma_N)^{-1}\|_{\infty}\|I_N-\Gamma_N\|_{\infty}\\
%&\leq2\cosh(\|\vv{Q}\|_1)\left(1+\|\mathcal{K}\|^2_{\infty}\|\vv{G}\|^2_1\right),
%\end{split}
%\end{equation}

Turning to the solution of the linear 
system $\left(I_N-\Gamma_N\right)\vv{A}_N=(\vv{E}_0)_N$, there are two options, namely, the 
direct method (possibly a sparse solver which takes into account the sparsity of $\Gamma_N$) or an 
iterative solver such as the BiCGSTAB~\cite{V1992}, a stable variant of the biconjugate 
gradient method. For each value of the spectral parameter, the complexity of computing 
the scattering coefficient using the direct sparse solver is less than $\bigO{N^3}$ 
while the same for the sparse iterative solver is less than $\bigO{N_{\text{iter.}}N^2}$. A fast variant 
of the iterative solver for $\mathcal{S}_N(\vv{A}_N,\vv{B}_N)^{\tp}=((\vv{E}_0)_N,\vs{0}_N)^{\tp}$ can be 
easily developed by observing that the matrix--vector product
$\mathcal{K}_N\mathcal{M}_N[\vv{G}_M]\vv{A}_N$ can be computed with complexity $\bigO{N\log N}$ using 
the fast polynomial multiplication in Chebyshev basis~\cite{G2011} (which in turn relies on the FFT 
algorithm) yielding a complexity of $\bigO{N_{\text{iter.}}N\log N}$ for each value of the 
spectral parameter. We label this fast iterative solver as FBiCGSTAB. Let us note that the 
FBiCGSTAB does not need to actually create the matrix $\mathcal{M}_N[\vv{G}_M]$ which makes it 
efficient in terms of usage of memory. The tolerance for the 
iterative solvers is set to be $10^{-12}$ and the maximum iteration threshold to $50$.

\section{Numerical Tests}
In this section, we conduct numerical tests in order to confirm the convergence behavior 
and the complexity of the algorithms proposed. For this purpose we consider the 
chirped secant-hyperbolic potential~\cite{TVZ2004} given by $q(t)=Wf(t/W)$ for $t\in\field{I}$ where 
$f(s) = A_0{\exp[2i\mu A_0\log(\sech s)]}{\sech(s)}$ for $s\in\field{R}$. We restrict 
to the case $\alpha=-1$. The parameter $W>0$ controls how well the profile $q(t)$ is 
supported in $\field{I}$. Let $\mu \geq 1$ so that the discrete spectrum is empty. It can be 
shown that $|\mathfrak{b}(\xi)|\sim \text{const.}\times\exp[-\pi(|\xi/W|-\mu A_0)]$ for
$|\xi/W|\gg1$; therefore, the interval $[-\mu WA_0, \mu WA_0]$ can be considered
as the effective width of the $\mathfrak{b}$ coefficient. We set $\xi_{\text{max}}=\mu WA_0$ in the 
following. The numerical error is quantified by 
$e_{\text{rel.}}=\|\vv{f}-\vv{f}^{(\text{num.})}\|_{\ell^2}/\|\vv{f}\|_{\ell^2}$
where $\vv{f}$ comprises samples $f_n=\mathfrak{b}(\xi_n)\,\,\text{or}\,\,\rho(\xi_n)$ over some 
finite sequence $(\xi_n)$ and $\vv{f}^{(\text{num.})}$ denotes the numerically computed values.

\begin{figure}[h!]
\centering
\includegraphics[scale=1]{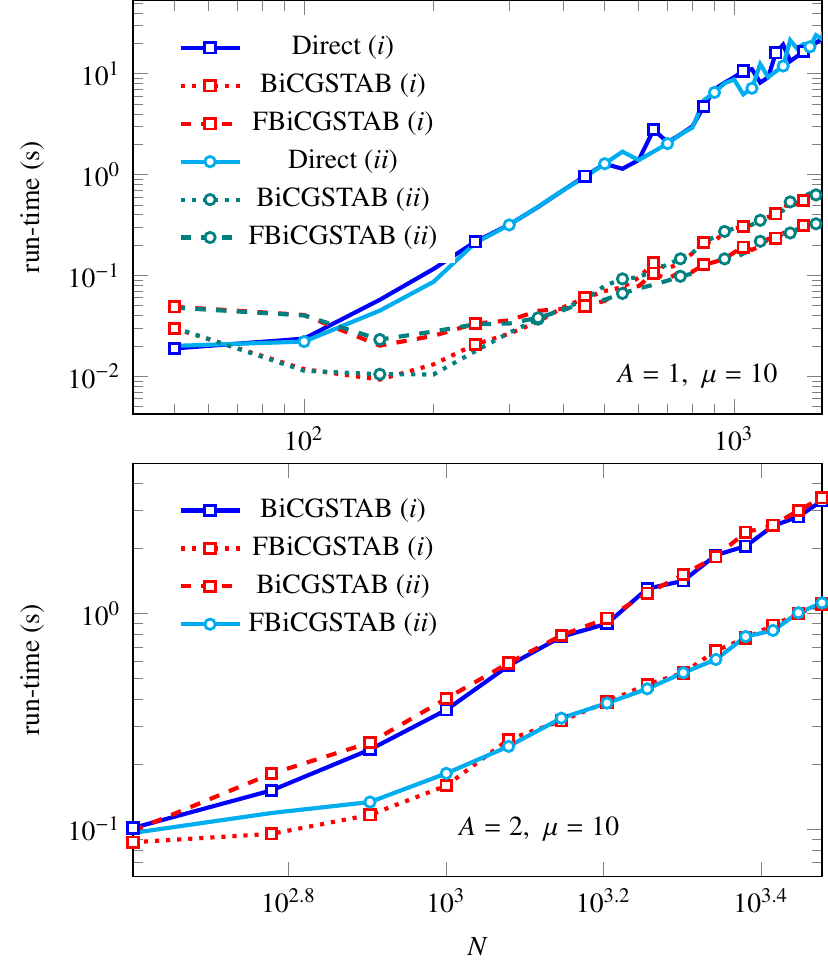}
\caption{\label{fig:runtime-all}Complexity analysis of 
the proposed algorithms for $(i)$ $\xi=\xi_{\text{max}}/2$, and, 
$(ii)$ $\xi=\xi_{\text{max}}$.}
\end{figure}

For a first test of convergence, we restrict ourselves to moderate values: 
$A_0\in\{1,2\}$, $\mu=10$ and $W=30$ with 
$M\leq3\times10^3$ and $N=4M$. The results of the convergence analysis for 
fixed $\xi\in\{\xi_{\text{max}}/2,\xi_{\text{max}}\}$ are shown in 
Fig.~\ref{fig:convg-all} which confirms the spectral convergence of the numerical 
scheme. Note that the direct solver has been dropped for the case $A_0=2$. The seed solution 
for the iterative algorithms were null vectors, however, the fast iterative
algorithm uses one final iteration of BiCGSTAB with the output of the former as the seed solution 
in order to improve the accuracy. This step can be omitted for large $M$ when forming 
$\Gamma_N$ becomes costly. The result of the 
complexity analysis is plotted in Fig.~\ref{fig:runtime-all} which shows that 
iterative solvers outperform the direct 
solver. Note that the iterative solvers become increasingly more
attractive when the samples of the continuous spectrum are needed on a spectral grid so 
that the solution of the previous grid point can be used as the seed solution for the current.
% \begin{figure}[h!]
% \centering
% \includegraphics[scale=1]{runtime_iter}
% \caption{\label{fig:runtime-iter}Complexity analysis of 
% the proposed iterative algorithms for $(i)$ $\xi=\xi_{\text{max}}/2$, and, 
% $(ii)$ $\xi=\xi_{\text{max}}$.}
% \end{figure}

\begin{figure}[h!]
\centering
\includegraphics[scale=1]{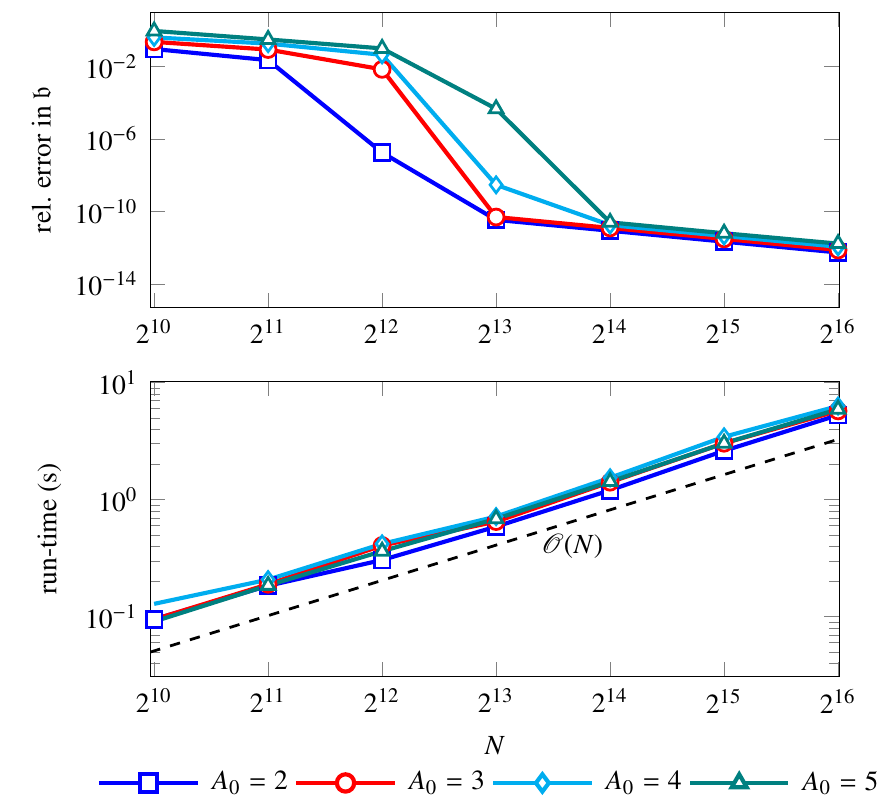}
\caption{\label{fig:result-fiter}Convergence (top) and complexity analysis for the fast method.}
\end{figure}

In the final test, we focus only on the fast method. Let $A_0\in\{2,3,4,5\}$ and 
$M\leq2^{16}$ ($N=4M$) while keeping $\mu=10$. In order to avoid early plateauing of 
error, we set $W=40$ because the chirped secant-hyperbolic profile 
does not strictly have a compact support. For the spectral parameter $\xi$, we choose a uniform grid over 
$[0,1.5\xi_{\text{max}}]$ of $20$ points. The result of the convergence analysis is 
plotted in Fig.~\ref{fig:result-fiter} (top) which once again confirms the spectral convergence 
of the numerical scheme. The complexity of the numerical scheme per spectral node can be 
estimated from Fig.~\ref{fig:result-fiter} (bottom) which confirms the quasilinear 
complexity. 

\begin{figure}[h!]
\centering
\includegraphics[scale=1]{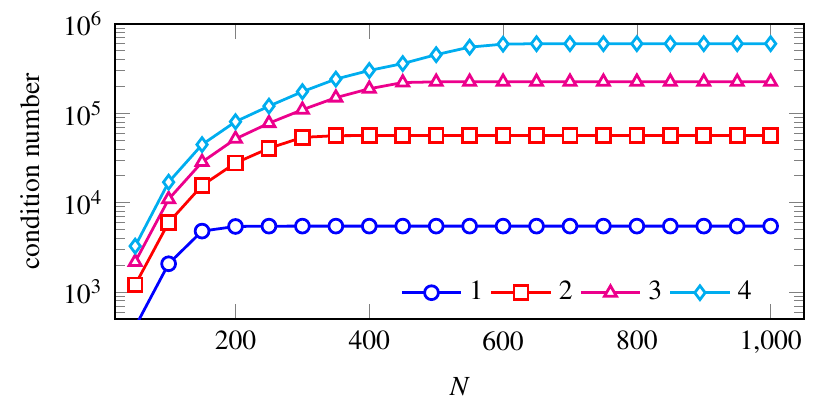}
\caption{\label{fig:cond-nr}The figure shows the condition number $\kappa(I_N-\Gamma_N)$ 
for $A_0\in\{1,2,3,4\}$ and $\mu=1.2$.}
\end{figure}

Let us conclude with the numerical evidence in Fig.~\ref{fig:cond-nr} 
for the claim that the condition number of $(I_N-\Gamma_N)$ remains bounded.

\bibliographystyle{IEEEtran}
%\bibliography{IEEEabrv,SNFT_PTL}

\providecommand{\noopsort}[1]{}\providecommand{\singleletter}[1]{#1}%

% Full bibliography added automatically for Optics Letters submissions
% Note that this extra page will not count against page length
%\ifthenelse{\equal{\journalref}{ol}}{%
%\clearpage
%\bibliographyfullrefs{NS_OL_fullref}
%\renewcommand{\refname}{Full References}
%}{}

\end{document}